# Event-Based Estimation for Detrusor Pressure Using a Single Bladder Catheter Augmented with Abdominal Electromyography

Mohamed Abdelhady, Reilly Burhanna, Lee Brody, Margot S. Damaser, *Senior Member*, *IEEE,* and Steve J.A. Majerus, *Senior Member*, *IEEE*

***Abstract*— *Objective:* The aim of this study was to enhance patient comfort and diagnostic accuracy in urodynamic studies (UDS) by developing a novel framework that uses a single vesical pressure (Pves) channel combined with a surface abdominal electromyography (AEMG) electrode array to estimate detrusor pressure (Pdet) without the need for an abdominal catheter. The proposed method was designed to operate in real-time, and to be compatible with current UDS clinical hardware. *Methods*: 40 UDS studies were performed with AEMG measured simultaneously with both Pves and Pabd. A predefined, provocative abdominal maneuver was used before each study to ensure accurate coupling between the AEMG signal and associated abdominal pressures seen in Pves. Criteria were set to determine if coupling was sufficient for Pdet estimation from Pves and AEMG. Post-hoc signal processing was used to further refine and evaluate estimated Pdet using automated, non-causal event detection. An event-based scoring metric was used to determine how accurately this approach estimated Pdet by calculating the percentage of signal contained in the appropriate channel for that event (Pdet vs abdominal pressure). *Results*: The results showed that 30 of the 40 cases indicated a "Go" status after the provocative maneuver, demonstrating feasibility of this method in a majority of UDS studies. The scoring metric, along with correlation analysis, confirmed accuracy of estimated Pdet with scores ranging from 85% to 95% with median at 93%. *Significance*: This innovative approach has the potential to revolutionize urodynamic studies by providing a more comfortable and accurate diagnostic tool for assessing bladder function. This is particularly beneficial for vulnerable populations, such as the elderly and children, reducing the need for multiple catheters and improving the overall patient experience during UDS.**

Manuscript submitted for review on April 15, 2025.

This work was supported in part by the US Department of Veterans Affairs, Case Western Reserve University, and SRS Medical. The contents do not represent the official views of the United States government or the United States Department of Veterans Affairs.

M. Abdelhady was previously with the Lerner Research Institute at the Cleveland Clinic.

M.S. Damaser is with the Department of Biomedical Engineering at the Cleveland Clinic Lerner Research Institute, with additional appointment in the Glickman Urological and Kidney Institute.

S.J.A. Majerus and R. Burhanna are with the Department of Electrical, Computer, and Systems Engineering, Case Western Reserve University.

M.S. Damaser and S.J.A. Majerus are with the Advanced Platform Technology Center, Louis Stokes Cleveland Veterans Affairs Medical Center, Cleveland, OH.

L. Brody is with SRS Medical, based in Chelmsford, MA.

## I. Introduction

Urinary incontinence (UI) is a widespread health issue affecting over 200 million people globally, significantly impacting their quality of life and placing a considerable burden on healthcare systems worldwide [1], [2]. The condition, defined by the involuntary loss of urine, arises from various underlying mechanisms such as detrusor overactivity (DO), stress urinary incontinence (SUI), and detrusor underactivity (DU). Accurate diagnosis of lower urinary tract dysfunctions (LUTD) that manifest as UI is essential for effective treatment [3].

Diagnostic methods range from clinical assessments using patient history and physical examinations to more intricate evaluations such as urodynamic studies (UDS), which can be crucial for developing tailored and effective bladder management strategies [2]. UDS is a collective term for a variety of diagnostic tests used to assess the function of the lower urinary tract [2]. The goal of UDS is to reproduce the patient's symptoms and provide a pathophysiological explanation by identifying factors that contribute to LUTD, including those that are asymptomatic.

Detrusor pressure (Pdet), the contractile force of the bladder muscle, is investigated during the cystometry portion of UDS. A transurethral catheter is inserted into the bladder through the urethra to fill the bladder with sterile fluid and measure the pressure inside the bladder (vesical pressure, Pves) during both rapid retrograde filling of the bladder and voiding around the catheter. However, Pves is influenced externally by abdominal pressure (Pabd), which is exerted by the abdominal muscles on internal organs, including the bladder. Therefore, another catheter is inserted into the rectum or vagina to measure Pabd during the examination. Pdet is then calculated as the simultaneous subtraction of Pabd from Pves:

$$Pdet = Pves - Pabd \quad (1)$$

Pdet serves as a critical tool for isolating and diagnosing bladder contraction events from abdominal-induced pressures resulting from coughing, laughing, or shifts in posture [1], [2], [4]. However, the Pabd signal is error prone, especially when the coupling between Pabd and Pves is not perfect [5]. Previous work demonstrated the feasibility of estimating Pdet from Pves, relying on statistical inference and wavelet multiresolution analysis to isolate abdominal artifacts in Pves [6], [7]. While effective in removing transient artifacts from Pves, this method did not provide precise estimations of Pdet during Valsalva maneuvers or abdominal pushing.

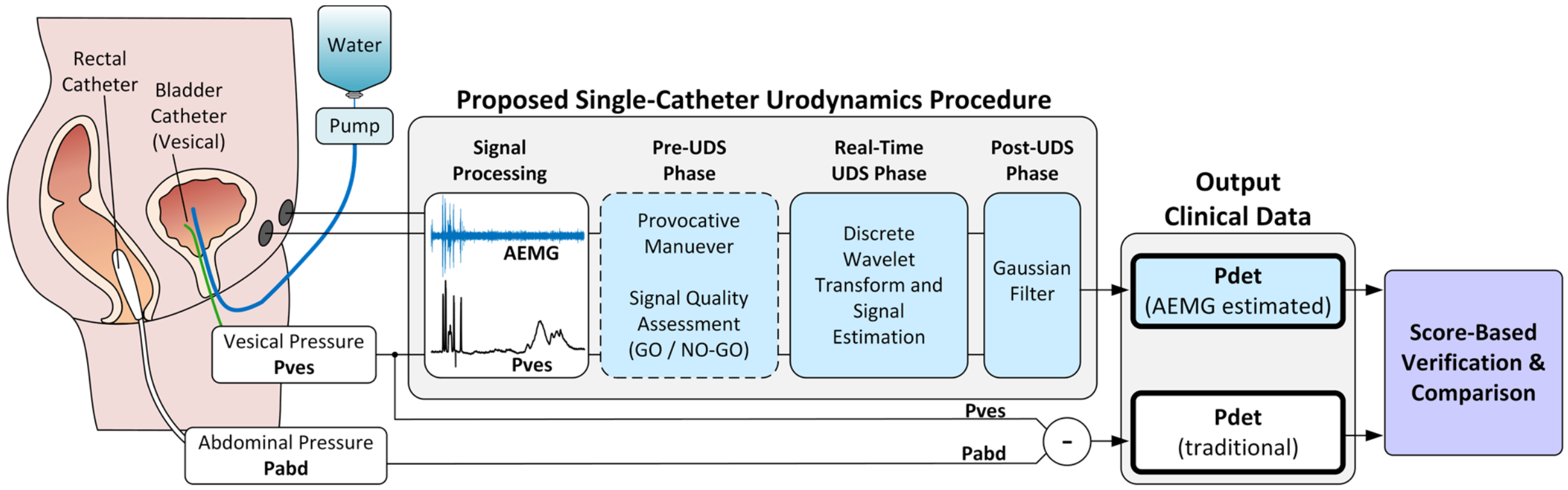

**Figure 1**. Detrusor pressure (Pdet) estimation included vesical (Pves) and surface abdominal electromyography (AEMG) signals processed in real-time during the UDS phase. A pre-UDS phase verifies proper correlation between AEMG and abdominal pressure and calculates the delay between AEMG and pressures in the bladder. To validate performance against traditionally calculated Pdet (using measured abdominal pressure), a score-based approach was used as detailed in this work.

Additionally, interpreting combined events, such as abdominal pushing during voiding proved challenging [7].

Abdominal electromyography (AEMG) presents a non-invasive method to estimate Pabd during UDS, eliminating the need for an abdominal catheter to measure Pabd [8]. This study aimed to integrate surface AEMG into UDS, potentially eliminating the need for an abdominal catheter (Fig. 1). This would improve patient comfort, reduce catheter position-related artifacts, and halve the number of disposable catheters used.

## II. Data Collection & Pre-Processing

### A. Overview

The UDS analysis for this study was done retrospectively but, to better match clinical expectations for real time data, was designed for real-time clinical use through three separate stages: pre-UDS, clinical-UDS, and post-UDS. The pre-UDS stage validated the signal integrity of AEMG and Pves data through a provocative maneuver and a “Go-No-Go” selection process. This preliminary data was additionally used to calculate patient specific tuning values for the later stages of the study. In the clinical UDS stage, Pves and AMEG data were processed in real-time to estimate detrusor pressure (Pdet) based on a discrete wavelet transform analysis. The post-UDS stage allowed for more computationally intensive filtering of the estimated Pdet signal before comparison to the traditionally calculated method shown in equation (1).

### B. Data Collection

This study was conducted in accordance with ethical standards and was determined to be exempt from IRB review by Sterling IRB (IRB ID: 9625-JBoczko, Exemption Category 4 per 45 C.F.R. §46.104(d); determination date: January 20, 2022). Data were collected from consecutive UDS studies at two US urology sites on adult male and female patients who presented with LUTD. Cystometry UDS recordings (n=40) were prospectively collected using 7 French air-charged catheters (TDOC-7F series, Laborie Medical Technologies, Portsmouth, NH) with an EasyPro Urodynamics system (SRS Medical, Chelmsford, MA) at 1,000 samples/sec. The 40 subjects recruited for this study consisted of 11 men and 29 women with demographic details typical for the LUTD population (Table I).

TABLE I: Patient Demographic Summary for 40 UDS Studies

| Age | Years | Gender | Count | Percentage |
|---|---|---|---|---|
| Range [min, max] | [41, 88] | Male | 11 | 27.5 |
| Median | 73 | Female | 29 | 72.5 |
| Mean (±SD) | 70.6 ± 12.2 | Total | 40 | 100 |

UDS recordings consisted of multiple signals including infused volume, Pves, and Pabd measured via a rectal catheter. Detrusor pressure (Pdet) was calculated using the simultaneous difference between the two pressure channels (Eq. 1). A surface AEMG electrode array (Smart Sensor series, SRS Medical) was placed approximately two inches both laterally and below the patient naval. Similar to the EasyPro measurements, AEMG data were sampled at 1,000 Hz. Recordings were saved digitally and processed retrospectively in MATLAB (Fig. 1).

### C. Pves and AEMG Signal Processing

The preprocessing stage was designed to filter and clean artifacts in the provocative maneuver data resulting from patient movements (Fig. 2). AEMG prefiltering began after a buffer of 4,000 samples was collected. Buffered data were downsampled without aliasing by one order of magnitude to reduce computational load while preserving essential information. Following downsampling, a 10 Hz low-pass filter (LPF) was employed to preserve clinically relevant, physiological bladder changes [9]. AEMG data were then rectified and the envelope of the rectified signal was detected using a moving root mean square (RMS) with a window length of 20 samples, corresponding to a history of 200 ms after accounting for the prior decimation. A length-50 moving average filter further smoothed the AEMG envelope [8].

Preprocessing of Pves data was optimized to reject large transients without removing Pdet contributions. While the initial steps of down-sampling and buffering mirrored those of AEMG data, the prefiltering block for Pves incorporated a two-stage filtering process beginning with a $10^{th}$-order finite impulse response (FIR) 10-200 Hz band-pass filter with 30 dB stopband attenuation. The second stage employed a 3rd order Savitzky-Golay smoother [10].

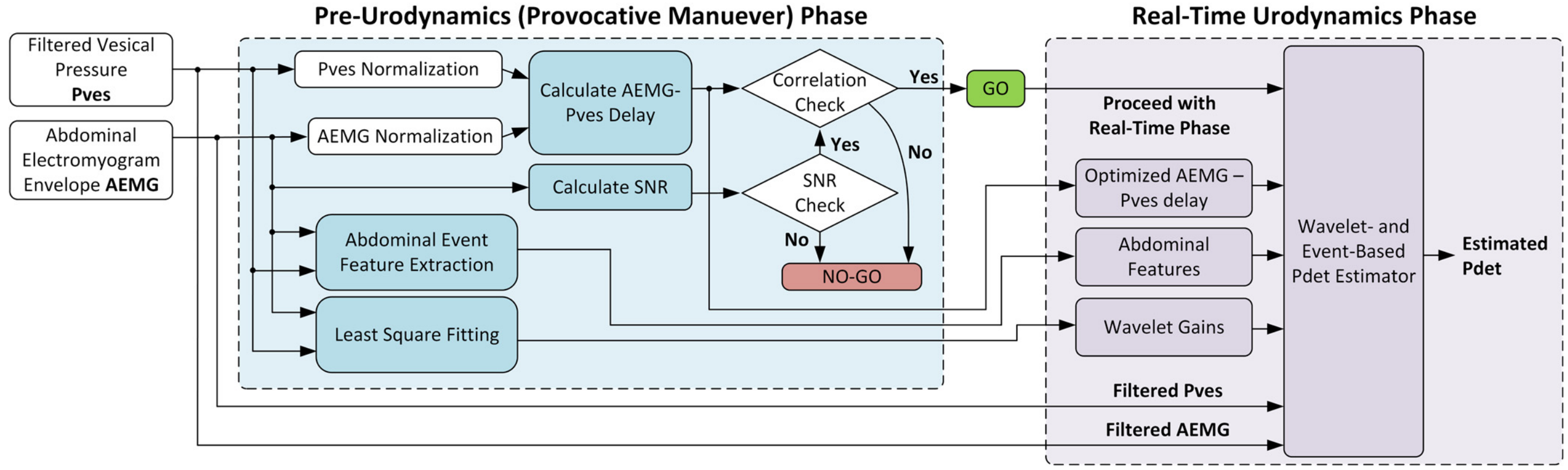

**Figure 2.** Processing pipeline of vesical pressure and abdominal electromyography signals included both a pre-urodynamics and real-time phase. The pre-urodynamics phase ensured the patient performed maneuvers to inject abdominal pressures into the Pves signal, to be quantified and correlated against AEMG data. Specific estimation values (AEMG-Pves delay), abdominal feature prominence and duration, and wavelet reconstruction gains were extracted from the pre-urodynamics phase and then used for the corresponding real-time phase in which bladder behavior is clinically observed.

### *D. AEMG Filtering Performance*

The AEMG pre-filtering stage was tested using 40 UDS recordings to determine suitability for power-line rejection and artifact removal to improve SNR of the detected AEMG envelope (Fig. 3). In 34 of 40 cases the AEMG SNR was above 48 dB, which was more than suitable for further estimation of abdominal pressure (Pabd). In six cases, AEMG SNR was in the range of 32-48 dB, which did not perform suitably well for a full analysis. Low AEMG SNR in these cases was estimated to be caused by larger body fat percentage and/or non-optimal placement of AEMG electrodes.

## III. Pre-Urodynamics Phase

### *A. Provocative Maneuver*

After the placement of catheters and AEMG and prior to the start of the bladder filling phase of the UDS study, the participants underwent a provocative maneuver procedure. This maneuver used voluntary coughs and Valsalvas (pushes) to validate the accuracy of AEMG and Pves data before beginning the UDS phase. Participants were instructed to execute a sequence of coughs and pushes designated as cough-push-cough. Ultimately, this approach was agnostic to the order or number of events in the provocative maneuver but relied only on pressure changes in Pves arising from abdominal pressures. Data from the provocative maneuver phase were used to determine suitability of the AEMG electrodes to estimate Pabd, through a “Go, No-Go” selection process (Fig. 4).

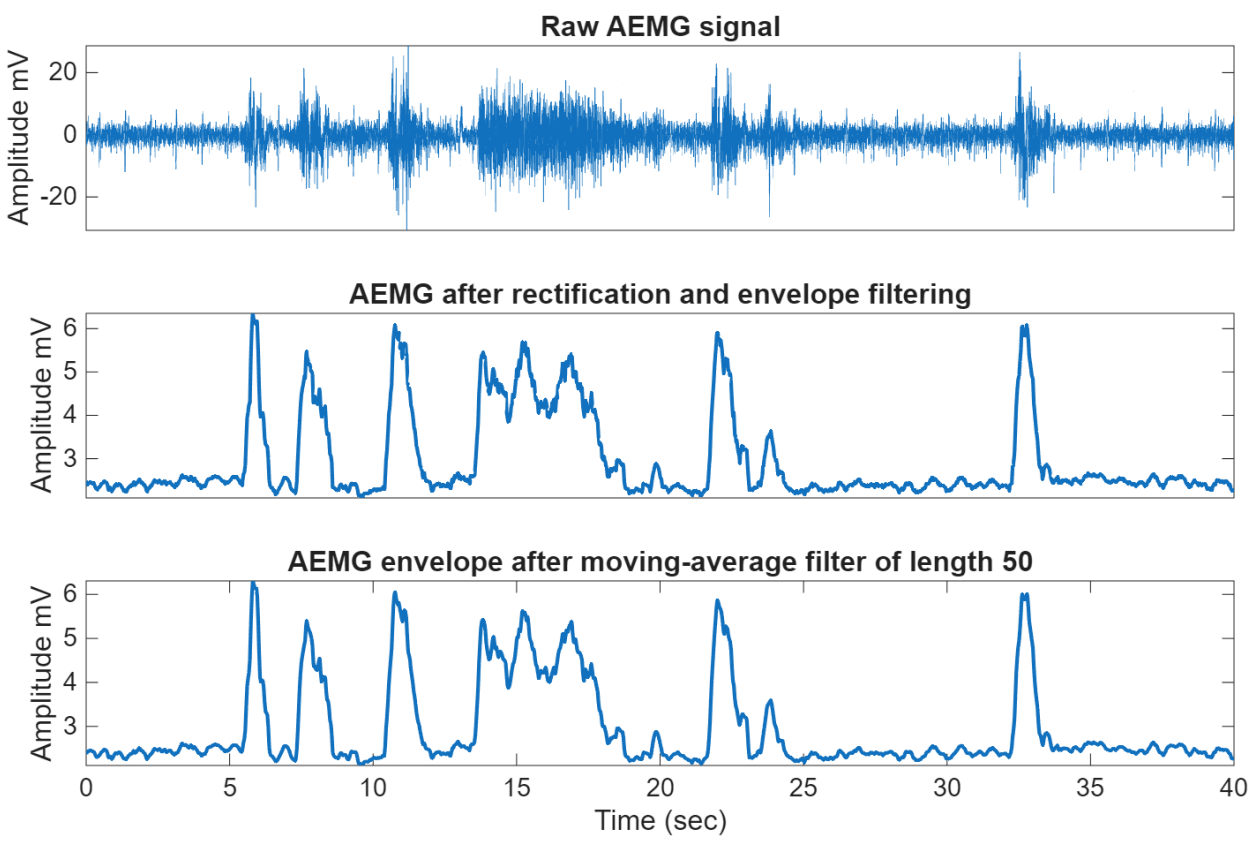

**Figure 3**. The raw measured AEMG signal was rectified, envelope-detected, and smoothed using a length-50 moving average filter. The resulting AEMG envelope was used as a surrogate measure of abdominal pressure (Pabd) traditionally measured using a catheter in UDS.

#### *1) AEMG and Pves Channel Quality Validation (Go-No-Go)*

Provocative maneuver data were assessed to ensure adequate coupling between AEMG and abdominal pressures. This process required a minimum Signal-to-Noise Ratio (SNR) of the AEMG channel, and sufficient AEMG-Pves similarity. AEMG-Pves similarity was calculated using coherence, computed as the Euclidean distance $d = \|v_p - v_e\|_2$ between the AEMG and Pves vectors. A smaller Euclidean distance therefore indicated greater similarity. To achieve this, we defined a threshold for the signals ($Te$ and $Tp$, with values calculated from the provocative maneuver) to optimize the similarity assessment.

#### *2) Go-No-Go Effectiveness*

In our study of 40 UDS recordings, 10 were classified as "No-Go" cases. Six of the No-Go cases had a Signal-to-Noise Ratio (SNR) below 40 ± 8 dB, indicating inadequate signal clarity for effective analysis. The remaining four cases in the No-Go category were attributed to vesical catheter movement during provocative maneuvers, which caused abrupt data surges and compromised the accuracy of the readings. The lowest acceptable SNR was determined to be -1.4 dB before a loss of correlation between Pves and AEMG.

The best results occurred when the AEMG threshold ($T_e$) was set at 20% of the maximum normalized AEMG value, and the Pves threshold ($T_p$) was set at 50% of the maximum normalized Pves value. This combination resulted in the smallest Euclidean distance of 0.1 ± 0.1, indicating the closest similarity between the AEMG and Pves channels. The average Euclidean distance, for the 30 successful UDS studies, was 0.6 ± 0.01.

The insights gained from these provocative maneuvers were instrumental in developing multistage filtering methods for both Pves and AEMG data. AEMG filtering led to an overall attenuation of 25 dB requiring that AEMG data be treated as a normalized quantity relative to maximum AEMG observed during the provocative maneuver. In contrast, Pves filtering resulted in a modest attenuation of 5 dB, which contributed to an offset-free detrusor pressure profile.

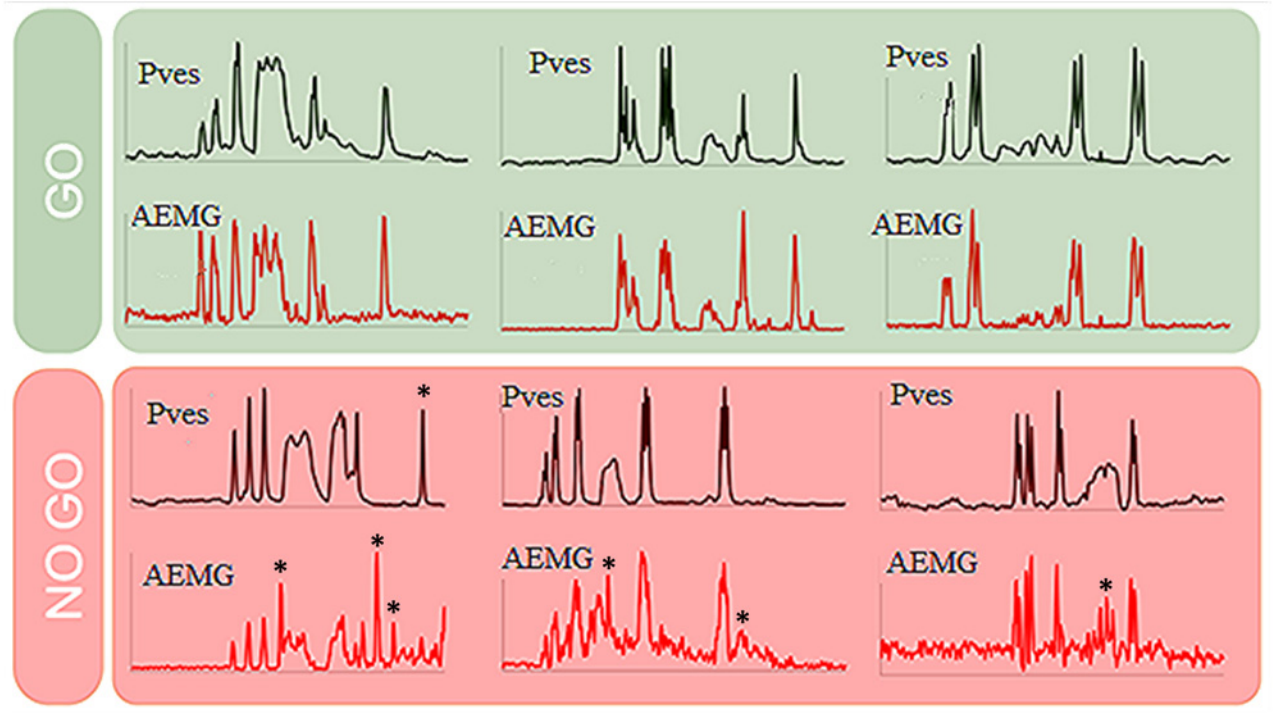

**Figure 4.** Examples of Go and No-Go cases in 6 patients, based on measured signal quality in a provocative maneuver. No-Go cases resulted from low correlation between AEMG and Pves channels, or low SNR. Asterisk symbols indicate uncorrelated events.

*B. Patient-Specific Tuning*

A delay averaging 1.8 ± 0.4 seconds between AEMG events and corresponding Pves events was identified during the preprocessing stage, predominantly due to the neuromechanical delay of muscle activation (Fig. 5) [11]. To detect and eliminate the delay between AEMG and Pves signals during the provocative maneuver, a variable-delay cross-correlation was computed. The delay calculation was performed just after the provocative maneuver, in which AEMG and Pves were assumed to correlate strongly to abdominal pressures (Fig. 5).

The lag at which the cross-correlation peaked was used as the time delay between Pves and AEMG. The cross-correlation function $R_{xy}(k)$ was calculated as:

$$R_{xy} = \sum_{n=0}^{N-1} \text{Pves}\,(n) \cdot \text{AEMG}(n+k) \tag{2}$$

where $N$ is total number of provocative maneuver samples, and $k$ is the lag. The cross-correlation peak was different for all patients, so this patient-specific delay value was used to align AEMG and Pves signals in subsequent processing. Fig. 6 illustrates a provocative maneuver extracted from a case study, demonstrating the effectiveness of removing the delay between AEMG and Pves signals by considering the Pves signal as stationary.

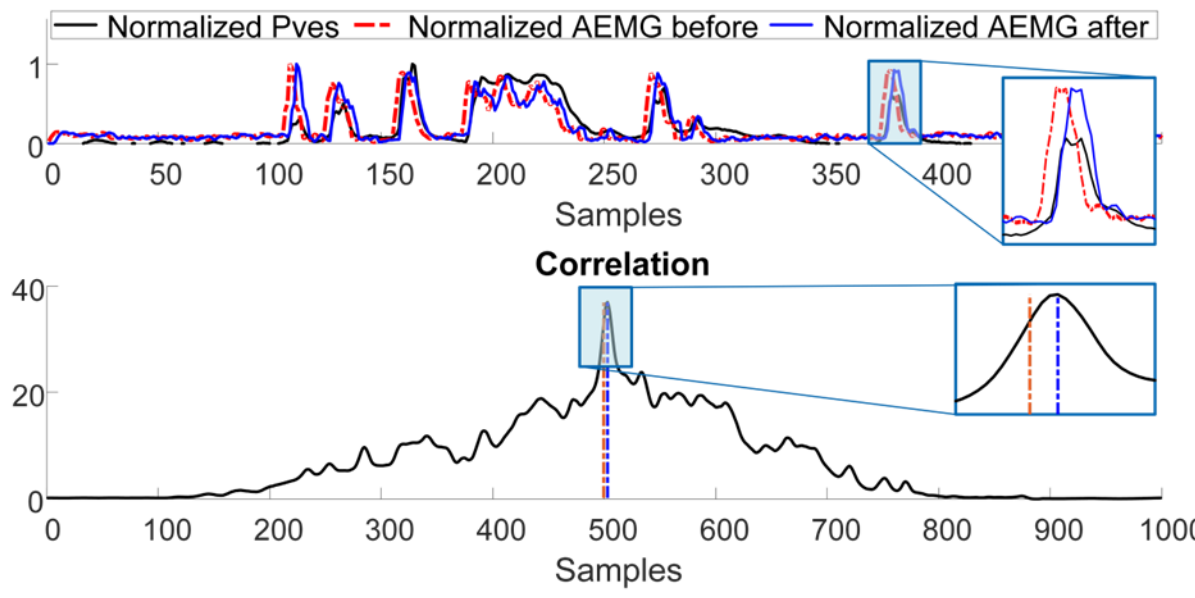

**Figure 5**. Example showing how the delay between maximum correlation (in samples) was used to offset the AEMG signal for time-alignment with pressure events in Pves. Data are normalized (unitless).

The features of the AEMG and Pves signals were characterized by analyzing the average values over each cough and push duration, and the rise and fall slopes of each of these maneuvers. Specifically, a temporal vector was created to record the timestamps of each cough and push maneuver separately for both channels. To identify these peaks, we considered the duration of the provocative maneuver as $L$ samples long. If the maximum number of potential coughs or pushes within this duration was $L/8$, we extracted $L/8$ peaks from each channel. These peaks were then filtered by selecting only those that exceeded a threshold for each channel.

Pves and AEMG signals from the provocative maneuver were normalized to the range of [0 1] so that detection thresholds $T_p$ for Pves events and $T_e$ for AEMG events could be set independently of patient-specific ranges (Fig. 6). Normalization simply divided the signal by the maximum value from the provocative maneuver:

$$\begin{aligned} Pves_{norm} &= {Pves}/{\max(Pves)} \\ AEMG_{norm} &= {AEMG}/{\max(AEMG)} \end{aligned} \tag{3}$$

To interpret these comparisons, we computed a performance index $D$ as the difference between the number of AEMG peaks exceeding the $T_e$ and the number of Pves peaks exceeding $T_p$:

$$D = \sum_{1}^{L} \left(EMG > T_e - Pves > T_p\right). \tag{4}$$

This ensured that significant events were accurately captured while accounting for variability in the provocative maneuvers. Finally, the timestamps associated with the AEMG and Pves were formed into two vectors as:

$$\begin{aligned} v_p &= [t_{c1}, t_{c2}, t_{cp1}, \ldots, t_{cn}, w_{c1}, w_{c2}, \ldots, w_{p1}, w_{cn}] \\ v_e &= [\bar{t}_{c1}\ \bar{t}_{c2}\ \bar{t}_{cp1} \ldots\ \bar{t}_{cn}, \bar{w}_{c1}, \bar{w}_{c2}, \ldots, \bar{w}_{p1}, \bar{w}_{cn}] \end{aligned} \tag{5}$$

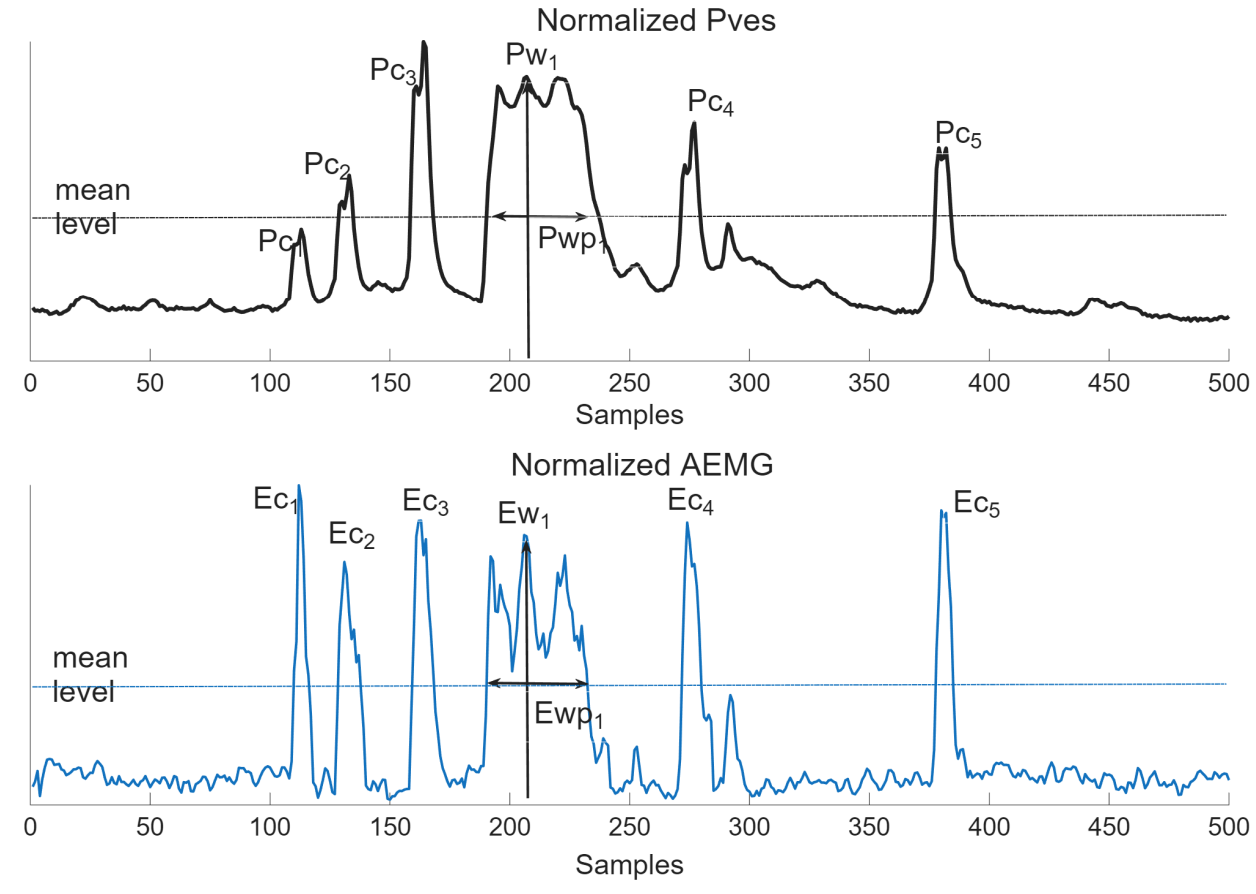

**Figure 6**. Temporal features of Pves and AEMG during an example provocative maneuver. Peak locations for coughs are indicated in both traces, while event duration is indicated for the Valsalva maneuver.

where $t_{ci}$ , $t_{pi}$, $w_{ci}$, and $w_{pi}$ denote the timestamps associated with the *i-th* cough and push, and the cough and push widths respectively. $v_p$ and $v_e$ represent the feature vectors for Pves and AEMG channels respectively and the bar indicates AEMG parameters. The cough or push width was calculated by identifying the points where the signal crossed the threshold level ($T_e$ and $T_p$) (Fig. 6). Event onsets were detected when a cough or push signal amplitude crossed each threshold.

The preprocessing concluded with the calculation of wavelet level gains using the least mean square method to determine the weight of each signal decomposition post-application of the wavelet decomposition technique [12], [13]. These gains were retained for subsequent real-time estimation.

## IV. Clinical Urodynamics Phase

### A. Detrusor Estimation Scheme

After successfully passing the Go-No-Go decision point, data were continuously processed as if in real time. A wavelet multiresolution analyzer (MRA) split incoming data into fundamental frequency components. The MRA used a tunable weight vector $W_i$ , where $i \in [1,2, \ldots N]$ and $N = log_2(W_{sz})$ as the wavelet resolution level. The weight vector allowed for estimating Pdet using a weighted sum. With $W_{sz}$=320, the 320-sample frame was decomposed into five approximation levels, as shown to be effective for bladder data [14].

The final level of the transform yields a single approximation coefficient due to decimation at each stage of the transform. We selected the Symlet wave function with four vanishing points (Sym4) for the MRA, based on a reconstruction analysis previously described [12], [13].

The output of the wavelet decomposer, $\tilde{P}$, is:

$$\bar{P}(W, \psi, W_{sz}) = \sum_{i=1}^{N} \left(\frac{1}{\sqrt{2}}\right)^{i} W_i \psi_i \tag{6}$$

where $W_i$ is the window resolution at level $i$, $\psi_i$ is the mother wavelet function, and $N = \log_2(W_{sz}) \in \mathbb{R}$ is the depth of decomposition. Vector $W_{sz}$ is predefined and selected to capture the slowest event of interest. In our work the slowest event is a push which has an average duration of $W_{sz} = 3.2$ ±0.3 sec.

Event detection enhances the DWT estimate of Pdet by enabling event-specific selection of the reconstruction weights $W_i$. For instance, when an abdominal event containing high frequency components is detected, the weighting vector applied to windows during the event can down-weight the initial DWT scales. The reconstruction weights $W_i$ are computed adaptively, allowing them to reflect recent activity in the data and thereby support a more robust detection framework. In this approach, we focused on using statistical features to adaptively scale $W_i$.

TABLE II: Average Characteristics for Coughs and Pushes in 40 UDS Studies

| Feature | Cough | Push |
|---|---|---|
| Rise Rate (cm H2O/sec) | 1.87± 0.43 | 4.5 ± 1.51 |
| Fall Rate (cm H2O/sec) | -2.7± 0.91 | -0.7 ± 0.4 |
| Duration (sec) | 1.1± 0.44 | 2.5 ± 0.5 |
| Peak (cm H2O) | 22 ± 3.7 | 15.6 ± 5.2 |
| Baseline $MAD_0$ (cm H2O) | 2.3 ± 1.8 | – |
| Baseline SNR (%) | 0.72 ± 0.2 | – |

Local statistical features of each window provide an alternative abstract representation of the signal and serve as strong indicators of contraction onset and termination for each event. The event span is defined as the time interval between the onset and termination points. Accurately identifying this span is particularly important in single-channel UDS, as it allows the signal processing algorithm to return to the more accurate DWT-based Pdet estimation using approximation coefficients once the event has concluded. Data from the pre-Urodynamics phase showed that the average duration of a cough was 1.4±0.5 sec and the average duration of a push event was 2.5±0.5 sec. Table II shows these average characteristics of the coughs and pushes during the provocative maneuvers in 40 UDS studies. The fixed window size of 320 samples (3.2 sec) was sufficient to capture the slowest event with at most two consecutive windows.

The event span cannot be calculated in real-time because the estimation process uses more than one sample. To address this, a flag signal $F_g$ is used. This flag is instantly turned on when triggered by any rise rate shown in Table II. The flag remains on until the fall rate condition meets any of those specified in Table II, at which point the flag is cleared. The resulting signal often exhibits logical redundancy caused by encountering both fall and rise rates typical of coughs and pushes. Consequently, a retrospective analysis is performed every $\tau$ time units to rectify this signal. This time interval was chosen to be a multiple of the sliding window size.

Given normalized AEMG and Pves in a sliding window, the kernel of the event detection is

$$\Gamma(\tau) = \begin{cases} \mathbf{1} & if \left(\sum_{i=1}^{\tau} F_g > T_m\right) \cap \left( \left\|(f_{pves} - f_{emg})\right\|_2 < T_r\right), \\ \mathbf{0} & otherwise \end{cases} \tag{7}$$

where $f = [\mu, \sigma, \xi, \delta] \in \mathbb{R}^4$ is the local statistical feature vector, $\sigma$, $\mu$, and $\delta$ are local standard deviation, arithmetic mean and maximum value of the local gradient, $\xi$ is the rate of signal zero crossing, $F_g$ is a flag set at onset and cleared at termination, and $T_m$ and $T_r$ are tuning parameters.

The baseline level of Pves before the event occurs is dynamically determined using the mean absolute deviation (MAD) which is preferred for its robustness to outliers and straightforward calculation, making it ideal for real-time applications where quick and reliable measures of dispersion are necessary [15]. The baseline level was calculated as

$$B = \begin{cases} (\text{Pves} - \bar{P}_5) & if \ \text{MAD}(\text{Pves} - \bar{P}_5) < \text{MAD}_0 \\ 0 & otherwise \end{cases} \tag{8}$$

where $B$ represents the baseline value, $\bar{P}_5$ denotes the wavelet 5th decomposed level of Pves, and $MAD_0$ corresponds to the value illustrated in Table II.

Finally, estimated Pdet ($P_{det}^{est}$) was calculated by eliminating detected events based on the condition

$$P_{det}^{est} = \begin{cases} B & if \ \Gamma(k) = 1 \\ \bar{P}_5 - \mu & otherwise. \end{cases} \tag{9}$$

This simple technique was adopted for low computational overhead in a real-time implementation.

# V. Post-Urodynamics Phase

## A. Postprocessing

Further processing of the real-time data was performed after the UDS was completed. This post-processing of data allowed the use of non-causal filtering and more computationally intensive steps. While this could be employed during the real-time portion (with some additional delay), post-hoc Gaussian filtering was used to remove additional non-physiological transient pressure fluctuations for detrusor pressure (Pdet). The filter output *g(x)* was calculated per input sample,

$$g(x) = \frac{1}{\sqrt{2\pi\sigma}} e^{\frac{-(x-\mu)^2}{2\sigma^2}}, \tag{10}$$

where $\mu$ and $\sigma$ represented the local mean and standard deviation, respectively, of a small sliding window of data points. The Gaussian filter computed a weighted average of the estimated Pdet samples within the window, with the weights determined by the Gaussian distribution. This process effectively smoothed the Pdet signal within each window, reducing high-frequency noise and fluctuations while retaining the underlying temporal dynamics of bladder activity. After filtering a window, the process slides the window by one data point and repeats the filtering operation. This iterative approach continues until the entire Pdet signal is processed, resulting in a smoothed Pdet signal with reduced noise.

# VI. Accuracy Estimation and Scoring

The current clinical practice for determining detrusor pressure subtracts Pabd from Pves measured simultaneously, as shown in Eq. (1). Any imperfect coupling from Pabd to Pves, or artifacts caused by the abdominal catheter, therefore propagate into artifacts on the calculated Pdet signal [16]. Because the proposed AEMG-based approach does not have the same artifact modes, a direct head-to-head comparison between estimated detrusor pressure ($P_{det}^{est}$) against the traditional calculated signal ($P_{det}$) would not be representative of overall accuracy.

Instead, an event-based validation method was developed to allow comparison between $P_{det}^{est}$ and $P_{det}$ signals based on accurate reproduction of clinically relevant data (i.e. bladder contraction events). Periods of bladder inactivity (containing no events) were not scored as they are not clinically relevant. Major event types analyzed are detailed in Table III.

To compare the calculated and estimated detrusor pressure during events, we employed a score metric, *S*, calculated as

$$S = 100 \cdot \left(1 - \sqrt{\left(\frac{\psi_{det}}{\psi_{pves}}\right)}\right), \tag{11}$$

where $\psi_{det}$ and $\psi_{pves}$ were the mean square pressure of the event portion after removing the baseline offset for Pdet and Pves,

$$\psi = \frac{1}{N}\sum_{i=1}^{N}(P_i - B)^2. \tag{12}$$

This event-based metric provided a more nuanced assessment of estimation accuracy, particularly in the context of clinically significant events. For each UDS study, the event portions, excluding coughs and pushes during provocative maneuvers, were manually labeled and evaluated using the score metric $S$ to produce a percentage of pressure for each event that was represented in Pdet and Pabd channels. The average of these individual scores represented the total score for each UDS trial in the study.

Event-based estimated and calculated Pdet signals were further compared through a similarity percentage based on the normalized cross-correlation coefficient where x[n] and y[n] are $P_{det}^{est}$ and $P_{det}$ respectively:

$$\text{Similarity (\%)} = 100 \cdot \frac{\sum_{n=1}^{N}(x[n]-x)(y[n]-y)}{\sqrt{\sum_{n=1}^{N}(x[n]-x)^2}\sqrt{\sum_{n=1}^{N}(y[n]-y)^2}} \tag{13}$$

# VII. Case Study Example

The proposed estimation technique is provided through an example (Fig. 7) with 5 highlighted phases. Here, estimated Pdet, $P_{det}^{est}$, was calculated by subtracting $P_{abd}^{est}$ from Pves. This example begins with a provocative phase that lasted 20 seconds and had three coughs, one push, and two additional coughs. During this period, all recorded Pves data were attributable to abdominal contributions, as expected. The second phase included a subsequent series of coughs, and the third included a possible patient effort to initiate voiding (evidenced by large AEMG amplitude). An overlap between the third and fourth phases indicated an abdominal push followed by a bladder contraction (BC). The fifth phase included a cough occurring during bladder voiding.

TABLE III: Event description corresponding to abdominal and vesical channel activates

| Event Name | Description | Activities | |
|---|---|---|---|
| | | **Abdominal** | **Pdet** |
| Cough | Short abdominal contraction | Yes | No |
| Valsalva | Extended abdominal push | Yes | No |
| Artifact | Artifact on Pves with no abdominal or detrusor activity | No | No |
| Bladder contraction | Bladder contraction with or without void | No | Yes |
| Bladder contraction with abdominal contribution | Bladder contraction with abdominal push | Yes | Yes |
| Bladder compliance | Passive increase in bladder pressure during filling phase | No | Yes |
| Baseline (no event) | The concatenation of all periods in the study in which there are no events | No | No |

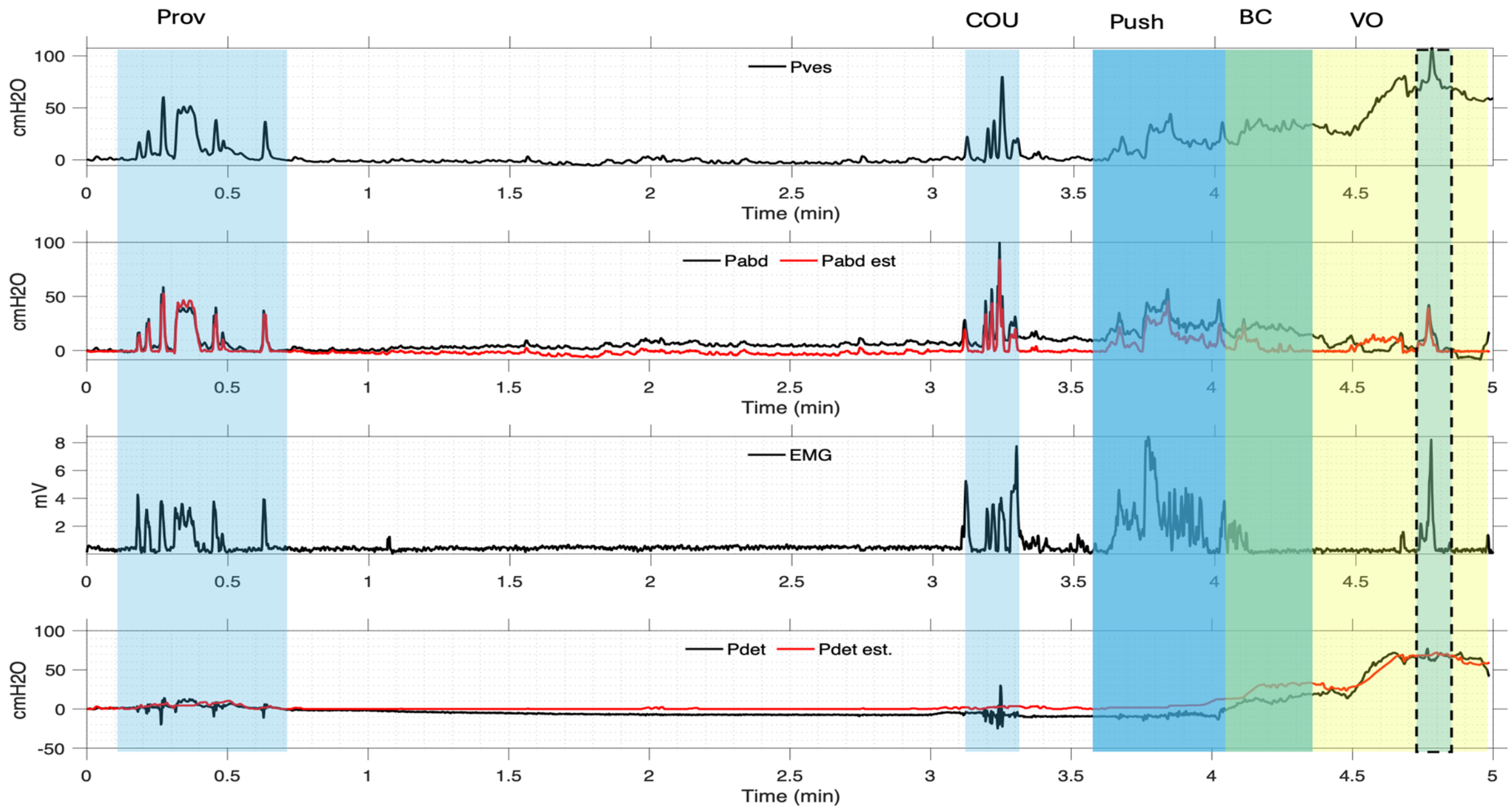

**Figure 7**: One example trace (out of 30) analyzed in this study, to compare accuracy between $P_{det}^{est}$ and traditionally $P_{det}$. The outcome of detrusor pressure (Pdet) estimation given the Pves (1st row) and surface AEMG (3rd row). The 2nd row depicts the measured abdominal pressure (Pabd) vs. $(Pves - P_{det}^{est})$. The bottom rows show $P_{det}$ vs. $P_{det}^{est}$. The shaded areas, from left to right, represent the provocative maneuver (Prov), cough (COU), push, bladder contraction (BC), and voiding (VO), including cough and the remainder of voiding.

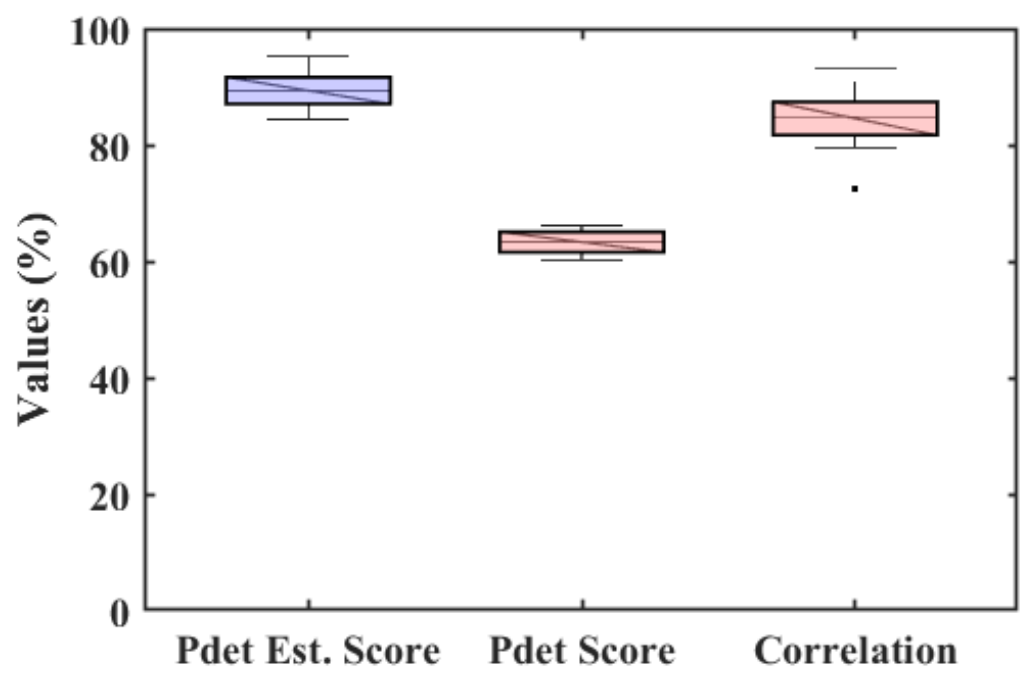

**Figure 8**: Estimation Score and Correlation were calculated for $P_{det}$ and $P_{det}^{est}$ across all recorded studies. $P_{det}^{est}$ achieved a median 93% Score [range: 85-95%]. The standard clinical $P_{det}$ achieved a median 64% Score [range: 30-93%], reflecting the presence of un-corrected artifacts arising from the abdominal catheter channel. Correlation between $P_{det}$ and $P_{det}^{est}$ had median value of 87% [range: 82-92%]. Box size represents the interquartile range (IQR) and whiskers are 1.5 times IQR.

## VIII. Accuracy Of Single-Catheter Approach

In most of the analyzed urodynamic studies, $P_{det}$ and $P_{det}^{est}$ showed strong alignment. However, $P_{det}$ occasionally exhibited errors resulting from subtraction, leading to physiologically impossible negative values due to misalignment of Pves and Pabd, particularly during cough events. A particular challenge arose in calculating the score for coughs occurring during voiding. Therefore, after the detrusor pressure component was removed, the event was scored as a standard cough. This issue underscored the rationale for adopting an event-based accuracy metric to assess the overall performance of the proposed UDS approach.

For the example study discussed in detail in the previous section, the average scores [calculated using (13)] for $P_{det}^{est}$ were 96.3±2.8% for coughs, 95.2±2.2% for pushes, and 87.6±1.4% for bladder contractions, yielding an overall average score of 93±2.6%. On the other hand, applying the same metric to the clinical $P_{det}$ yielded the following results: 93.5±3.1% for coughs, 67.5±2.4% for pushes, and 52.7±4.7% for bladder contractions, resulting in an overall average score of 71.2±3.7%. Note in this case the reduced accuracy was due to artifacts induced by the abdominal pressure catheter, which were not present when Pabd was estimated by AEMG.

Comparison between $P_{det}^{est}$ and $P_{det}$ was viewed qualitatively during refinement of calculation tuning parameters. Across the 30 "Go" studies included in this analysis, $P_{det}^{est}$ resulted in a median score of 93%, while $P_{det}$ had a median score of 64% (Fig. 8), indicating that the event-based estimation accurately captured the physiological profile of Pdet for the events analyzed (Table III). The similarity percentage between $P_{det}^{est}$ and $P_{det}$ during events had a median value of 87% indicating adequate capture of urological events without the abdominal catheter.

## IX. Discussion

UDS is the standard objective clinical method for diagnosing LUTD; however, it has significant drawbacks and is often unable to reproduce patient complaints [17], [18]. Bladder wall thickness, measurable via ultrasound, has been explored as a surrogate marker for detrusor function [19], [20]. Additionally, near-infrared spectroscopy is being investigated as a non-invasive method to assess bladder oxygenation and detrusor activity, and condom catheters may also be used to characterize bladder functional capacity [21], [22]. These technologies hold the potential to eliminate the need for catheterization, greatly improving patient experience, but most of them do not measure bladder pressure, the most important objective measure of bladder function. Wearable technologies have further advanced LUTD diagnostics. Devices using capacitive sensing and bioimpedance allow for continuous bladder volume monitoring, representing a step toward home-based urodynamic assessments [23], [24]. Advances in wireless catheter-free telemetric bladder pressure monitoring also show potential in advancing home based UDS [25], [26]. Despite these advances, challenges persist in standardizing non-invasive diagnostic techniques across diverse patient populations and ensuring their clinical implementation. Validation studies are needed to confirm the reliability and reproducibility of these methods in clinical practice, and cost-effectiveness analyses are essential for evaluating their economic feasibility [27], [28].

In this work, we introduced a novel approach to enhance UDS clinical methodology by implementing a single catheter augmented with AEMG. This hardware modification allowed us to develop a novel algorithm to estimate detrusor pressure, addressing the limitations of the classical method for calculating detrusor pressure. As far as the authors are aware, this is the first study to employ such an approach, aiming to reduce the need for dual catheters during UDS. A key component of this approach was the design and implementation of a provocative maneuver to validate the signals obtained from the vesical catheter and surface AEMG. This maneuver involved specific actions such as coughing and pushing, which provoked clear responses in the measured signals, allowing for robust evaluation. The signal processing pipeline included a novel mechanism for identifying signal quality and determining a "Go" status, indicating sufficient quality for further analysis. Results showed that 75% of cases met the "Go" criteria, demonstrating the feasibility of this approach in a significant majority of cases.

Previous work reported using an adaptive threshold that tracked baseline pressure and detected when the Pdet estimate exceeded this plus a fixed threshold [29]. More recent work investigated the detection and classification of bladder events using wavelet analysis [12], [13]. While these efforts represent significant steps toward accurately detecting bladder events, they do not focus on estimating the detrusor pressure itself. Our work builds on these foundations by not only improving event detection but also introducing a novel approach to estimate detrusor pressure, addressing a gap left by these earlier studies.

To assess the effectiveness of the signal estimation process, a scoring technique was developed to reflect the physiological relevance of the estimated Pdet. This evaluation method aimed at achieving fairness in scoring, which overcomes the misleading use of root mean square error (RMSE) as a metric. While RMSE enables accurate comparison between a new approach and a standard one. UDS data collected via two catheters from heterogenous patients, despite being the clinical standard, contains pressure artifacts and offsets that are well documented shortcomings of UDS [4], [5], [16]. Therefore, validating a new approach for its ability to reproduce artifactual data to improve RMSE is meaningless. The proposed scoring method, along with correlation analysis between calculated and estimated Pdet, demonstrated the method's effectiveness. High correlation values and scores indicated that estimated Pdet closely followed calculated Pdet pressure, reinforcing the reliability of the proposed method.

However, a downside of our scoring system was its reliance on manual annotation of the start and end of each event, which may have introduced inaccuracies or bias. Furthermore, we recognize the need for more robust fault detection algorithms to handle hardware failures during the estimation process, such as identifying and mitigating catheter or electrode movement caused by patient activity.

The primary limitation of our approach was its dependence on event occurrence rather than the relationship between AEMG and vesical pressure. This reliance on events may have caused misinterpretation due to the presence of false events. A prospective study of clinical decision making based on estimated Pdet is needed to determine if estimated Pdet is sufficiently accurate for clinical decisions.

## X. Conclusion

This study introduced a novel, less invasive approach to UDS, promising improved patient comfort and diagnostic accuracy in clinical settings. The method presented here was assessed retrospectively with UDS data prospectively recorded simultaneously with AEMG. Prior to clinical use, future studies should determine the accuracy and utility of this method when used in real-time during UDS examinations on a broad spectrum of patients.

## Acknowledgement and Conflict of Interest

SRS Medical funded this research (Dr. Damaser, PI) and licensed relevant IP from the Cleveland Clinic. All authors are inventors on that IP and have significant financial interest in the outcome of this project. Mr. Brody is employed by SRS Medical and has significant financial interests in this project. Dr. Abdelhady was a consultant to SRS Medical.

## References

[1] R. Voelker, “International Group Seeks to Dispel Incontinence ‘Taboo,’” *JAMA*, vol. 280, no. 11, pp. 951–953, Sep. 1998, doi: 10.1001/jama.280.11.951.

[2] A. J. Wein, “Diagnosis and treatment of the overactive bladder,” *Urology*, vol. 62, no. 5, Supplement 2, pp. 20–27, Nov. 2003, doi: 10.1016/j.urology.2003.09.008.

[3] J. Wang, L. Ren, X. Liu, J. Liu, and Q. Ling, “Underactive Bladder and Detrusor Underactivity: New Advances and Prospectives,” *Int. J. Mol. Sci.*, vol. 24, no. 21, p. 15517, Oct. 2023, doi: 10.3390/ijms242115517.

[4] D. W. Gould, A. C. Hsieh, and L. F. Tinckler, “The effect of posture on bladder pressure,” *J. Physiol.*, vol. 129, no. 3, pp. 448–453, Sep. 1955, doi: 10.1113/jphysiol.1955.sp005369.

[5] S. Hogan, A. Gammie, and P. Abrams, “Urodynamic features and artefacts,” *Neurourol. Urodyn.*, vol. 31, no. 7, pp. 1104–1117, 2012, doi: 10.1002/nau.22209.

[6] R. Karam *et al.*, “Real-Time Classification of Bladder Events for Effective Diagnosis and Treatment of Urinary Incontinence,” *IEEE Trans Biomed Eng*, vol. 63, no. 4, pp. 721–9, 2016, doi: 10.1109/tbme.2015.2469604.

[7] F. Zareen, Z. Ouyang, S. J. A. Majerus, T. M. Bruns, M. S. Damaser, and R. Karam, “Detrusor Pressure Estimation from Single-Channel Urodynamics,” in *2022 44th Annual International Conference of the*

*IEEE Engineering in Medicine & Biology Society (EMBC)*, Jul. 2022, pp. 3718–3722. doi: 10.1109/EMBC48229.2022.9871663.

[8] K. S. Kim, J. H. Seo, J. U. Kang, and C. G. Song, "Implementation of a Multi-functional Ambulatory Urodynamics Monitoring System Based on Newly Devised Abdominal Pressure Measurement," *J. Med. Syst.*, vol. 34, no. 6, pp. 1011–1021, Dec. 2010, doi: 10.1007/s10916-009-9318-1.

[9] A. Gammie *et al.*, "International continence society guidelines on urodynamic equipment performance," *Neurourol. Urodyn.*, vol. 33, no. 4, pp. 370–379, 2014, doi: 10.1002/nau.22546.

[10] P. Gans and J. B. Gill, "Examination of the Convolution Method for Numerical Smoothing and Differentiation of Spectroscopic Data in Theory and in Practice," *Appl. Spectrosc.*, vol. 37, no. 6, pp. 515–520, Nov. 1983, doi: 10.1366/0003702834634712.

[11] R. Krätschmer, M. Stingl, D. Holzer, and F. K. Paternoster, "The neuromechanical delay of the quadriceps shortens with increasing contraction intensity," *Sci. Rep.*, vol. 15, p. 25378, Jul. 2025, doi: 10.1038/s41598-025-10477-1.

[12] S. J. A. Majerus, M. Abdelhady, V. Abbaraju, J. Han, L. Brody, and M. Damaser, "Real-Time Wavelet Processing and Classifier Algorithms Enabling Single-Channel Diagnosis of Lower Urinary Tract Dysfunction," in *Machine Learning Applications in Medicine and Biology*, A. Ahmed and J. Picone, Eds., Cham: Springer Nature Switzerland, 2024, pp. 87–114. doi: 10.1007/978-3-031-51893-5_4.

[13] M. Abdelhady, J. Han, S. J. A. Majerus, L. Brody, and M. Damaser, "Detrusor Pressure Estimation from Single Channel Bladder Pressure Recordings," in *2022 IEEE Signal Processing in Medicine and Biology Symposium (SPMB)*, Dec. 2022, pp. 1–5. doi: 10.1109/SPMB55497.2022.10014843.

[14] R. Karam, S. J. A. Majerus, D. J. Bourbeau, M. S. Damaser, and S. Bhunia, "Tunable and Lightweight On-Chip Event Detection for Implantable Bladder Pressure Monitoring Devices," *IEEE Trans. Biomed. Circuits Syst.*, vol. 11, no. 6, pp. 1303–1312, 2017, doi: 10.1109/TBCAS.2017.2748981.

[15] S. Guo, "Advanced Statistical Analysis," in *Encyclopedia of Social Work*, 2013. doi: 10.1093/acrefore/9780199975839.013.840.

[16] E. Finazzi Agrò, D. Bianchi, and V. Iacovelli, "Pitfalls in Urodynamics," *Eur. Urol. Focus*, vol. 6, no. 5, pp. 820–822, Sep. 2020, doi: 10.1016/j.euf.2020.01.005.

[17] J. Pannek and P. Pieper, "Clinical usefulness of ambulatory urodynamics in the diagnosis and treatment of lower urinary tract dysfunction," *Scand. J. Urol. Nephrol.*, vol. 42, no. 5, pp. 428–432, 2008, doi: 10.1080/00365590802299056.

[18] K. D. Clement, H. Burden, K. Warren, M. C. M. Lapitan, M. I. Omar, and M. J. Drake, "Invasive urodynamic studies for the management of lower urinary tract symptoms (LUTS) in men with voiding dysfunction," *Cochrane Database Syst. Rev.*, vol. 2015, no. 4, p. CD011179, Apr. 2015, doi: 10.1002/14651858.CD011179.pub2.

[19] M. Z. Nasrabadi, H. Tabibi, M. Salmani, M. Torkashvand, and E. Zarepour, "A comprehensive survey on non-invasive wearable bladder volume monitoring systems," *Med. Biol. Eng. Comput.*, vol. 59, no. 7–8, pp. 1373–1402, Aug. 2021, doi: 10.1007/s11517-021-02395-x.

[20] F. F. Farag and J. P. Heesakkers, "Non-invasive techniques in the diagnosis of bladder storage disorders," *Neurourol. Urodyn.*, vol. 30, no. 8, pp. 1422–1428, 2011, doi: 10.1002/nau.21155.

[21] F. F. Farag, F. M. Martens, K. W. D'Hauwers, W. F. Feitz, and J. P. Heesakkers, "Near-infrared spectroscopy: a novel, noninvasive, diagnostic method for detrusor overactivity in patients with overactive bladder symptoms--a preliminary and experimental study," *Eur. Urol.*, vol. 59, no. 5, pp. 757–762, May 2011, doi: 10.1016/j.eururo.2010.12.032.

[22] R. van Mastrigt, J. J. M. Pel, J. W. N. C. H. F. Chung, and S. de Zeeuw, "Development and application of the condom catheter method for non-invasive measurement of bladder pressure," *Indian J. Urol. IJU J. Urol. Soc. India*, vol. 25, no. 1, pp. 99–104, Jan. 2009, doi: 10.4103/0970-1591.45546.

[23] H. Ozawa, T. Igarashi, K. Uematsu, T. Watanabe, and H. Kumon, "The future of urodynamics: Non-invasive ultrasound videourodynamics," *Int. J. Urol.*, vol. 17, no. 3, pp. 241–249, 2010, doi: 10.1111/j.1442-2042.2010.02447.x.

[24] J. J. Park, A. Kwon, J. Y. Park, S. R. Shim, and J. H. Kim, "Efficacy of Pelvic Floor Exercise for Post-prostatectomy Incontinence: Systematic Review and Meta-analysis," *Urology*, vol. 168, pp. 175–182, Oct. 2022, doi: 10.1016/j.urology.2022.04.023.

[25] B. T. Frainey *et al.*, "First in Human Subjects Testing of the UroMonitor: A Catheter-free Wireless Ambulatory Bladder Pressure Monitor," *J. Urol.*, vol. 210, no. 1, pp. 186–195, Jul. 2023, doi: 10.1097/JU.0000000000003451.

[26] M. D. Gross *et al.*, "Validation of a Wireless Catheter-Free Ambulatory Urodynamics Device in Women With Neurogenic Bladder," *Neurourol. Urodyn.*, vol. 0, no. 0, pp. 1–9, 2025, doi: 10.1002/nau.70172.

[27] M. U. Ali, K. N.-K. Fong, P. Kannan, U. M. Bello, and G. Kranz, "Effects of nonsurgical, minimally or noninvasive therapies for urinary incontinence due to neurogenic bladder: a systematic review and meta-analysis," *Ther. Adv. Chronic Dis.*, vol. 13, p. 20406223211063059, 2022, doi: 10.1177/20406223211063059.

[28] I. Clausen, L. G. W Tvedt, and T. Glott, "Measurement of Urinary Bladder Pressure: A Comparison of Methods," *Sensors*, vol. 18, no. 7, p. 2128, Jul. 2018, doi: 10.3390/s18072128.

[29] R. Karam *et al.*, "Real-time classification of bladder events for effective diagnosis and treatment of urinary incontinence," *IEEE Trans. Biomed. Eng.*, vol. 63, no. 4, pp. 721–729, 2016, doi: 10.1109/TBME.2015.2469604.